\documentclass{epl}
\usepackage{amssymb}
\usepackage{rotating,psfig,epsfig}
\usepackage{graphics,color}
\usepackage{amsmath}

\newcommand{\bpm}[1]{\mbox{\boldmath $\pm$}}

\renewcommand{\epsilon}{\varepsilon}
\shorttitle{Gap solitons in a Bragg superlattice}
\shortauthor{K. Yagasaki \etal}
\institute{
\inst{1}
 Department of Mechanical and Systems Engineering, Gifu University,
 Gifu 501-1193, Japan\\
\inst{2}
 Department of Interdisciplinary Studies, Faculty of Engineering,
 Tel Aviv University, Tel Aviv 69978, Israel\\
\inst{3}
 Department of Engineering Mathematics, University of Bristol,
 Bristol BS8 1TR, UK
} 

\begin{document}

\title{Gap solitons in Bragg gratings with a harmonic superlattice}
\author{Kazuyuki Yagasaki\inst{1} \and Ilya M. Merhasin\inst{2} \and Boris
A. Malomed\inst{2} \and Thomas Wagenknecht\inst{3} \and Alan R.
Champneys \inst{3} } \maketitle

\begin{abstract}
Solitons are studied in a model of a fiber Bragg grating (BG) whose local
reflectivity is subjected to periodic modulation. The superlattice opens an
infinite number of new bandgaps in the model's spectrum. Averaging and
numerical continuation methods show that each gap gives rise to gap solitons
(GSs), including asymmetric and double-humped ones, which are not present
without the superlattice.Computation of stability eigenvalues and direct
simulation reveal the existence of \emph{completely stable} families of
fundamental GSs filling the new gaps -- also at negative frequencies, where
the ordinary GSs are unstable. Moving stable GSs with positive and negative
effective mass are found too.
\end{abstract}



Bragg gratings (BGs) are distributed reflecting structures
produced by periodic variation of the refractive index of an
optical fiber or waveguide. Devices based on fiber gratings, such
as dispersion compensators, sensors and filters, are widely used
in optical systems \cite{Kashyap}. Gap solitons (GSs) in fiber
gratings are supported (in the temporal domain) by the balance
between BG-induced linear dispersion, which includes a bandgap in
the system's spectrum, and the Kerr nonlinearity of the fiber
material. Analytical solutions for BG solitons in the standard
model of the fiber grating, based on coupled-mode equations for
the right- and left-traveling waves, are well known
\cite{AW-CJ:89}. Solitons with positive frequency $\omega $ are
stable, while those with $\omega <0$ have an instability to
perturbations that is characterized by a complex growth rate
\cite{BPZ:98}.

Solitons in fiber gratings have been created experimentally, with
spatial and temporal widths of the order of a few mm and$~50$ ps,
respectively \cite{experiment}. Spatial GSs were observed in
photorefractive media with an induced photonic lattice
\cite{Segev}, and in waveguide arrays~\cite{Silberberg}. (Indeed,
the new families of GSs reported in the present work may be
realized in the spatial domain too, in addition to their
straightforward implementation as temporal solitons in fiber
gratings). Besides their relevance to optical media, GSs were also
predicted \cite{BEC} and created \cite{Oberthaler} in a
Bose-Einstein condensate (BEC) trapped in a periodic potential.

An issue of particular technological importance is the development
of methods for the control of BG solitons -- in particular, using
\textit{apodization} \cite{experiment,express,WMak}, i.e.~ gradual
variation of the grating's reflectivity along the fiber. In an
appropriately apodized BG, one can slow down solitons and,
eventually, bring them to a halt \cite{WMak}. Experimentally, it
has been demonstrated that apodization helps to couple solitons
into the grating \cite{experiment}. Moreover, technologies are
available that allow one to fabricate fiber gratings with
\emph{periodic} apodization, thus creating an effective
\textit{superlattice} built on top of the BG \cite{Russell}.

An asymptotic analysis of light propagation in such a \textit{superstructure
grating} was developed in Ref.~\cite{SSG}. It was shown that the
superstructure gives rise to extra gaps in the system's spectrum (``Rowland
ghost gaps"). Solitons in the gaps were sought by assuming that the soliton
is a slowly varying envelope of the superlattice's Bloch function, which
applies to the description of GSs near bandgap edges. Related problems were
considered in Refs.~\cite{Louis}, which treat BEC models with a doubly
periodic optical lattice that opens up an additional narrow ``mini-gap", in
which stable solitons may be found.

The subject of this Letter is the investigation of GSs in harmonic
(sinusoidal) superlattices created on top of the ordinary BG in fibers with
Kerr nonlinearity. The analysis of the bandgap structure in this model
constitutes, by itself, an interesting extension of the classical spectral
theory for the Mathieu equation \cite{Mathieu}. We will demonstrate that
various families of GSs exist in the superlattice. Most importantly, in each
newly opened gap we find a family of fundamental symmetric GSs, which fill
the entire gap and are \emph{completely stable}. Remarkably, they are stable
not only at positive frequencies $\omega $, but also at $\omega <0$, where
the ordinary GSs are unstable \cite{BPZ:98}.

The superlattice may be implemented via the creation of beatings in the
optical interference pattern that burns the grating into the fiber's
cladding. A model corresponding to this physical situation may be derived
using standard coupled-mode equations for the amplitudes $u$ and $v$ of the
right- and left-traveling electromagnetic waves. Inclusion of both Kerr
nonlinearity and Bragg reflection terms leads to the following dimensionless
equations:
\begin{equation}
\begin{array}{rcl}
iu_{t}+iu_{x}+\left[ 1-\epsilon \cos \left( kx\right) \right] v+\mu \cos
\left( kx+\delta \right) ~u+\left( |v|^{2}+|u|^{2}/2\right) u & = & 0, \\
iv_{t}-iv_{x}+\left[ 1-\epsilon \cos \left( kx\right) \right]
u+\mu \cos \left( kx+\delta \right) ~v+\left(
|u|^{2}+|v|^{2}/2\right) v & = & 0.\end{array} \label{e:uv}
\end{equation}
Here $\epsilon $ is the amplitude of the periodic modulation of the BG
strength, while the nonlinearity coefficient and average
reflectivity are normalized, as usual, to be $1$. Accordingly, $k$
measures the ratio of the modulation period $L$ to the BG
reflection length $l$ (usually, $L$ $\lesssim 1$ cm
\cite{Russell,SSG} and $l\sim 1$ mm, while the underlying BG
period is $\sim 1$ $\mu $m, whereas the total length of the
grating may be up to $1$ m). We have included additional terms in
Eqs.~(\ref{e:uv}) with amplitude $\mu $ and phase shift $\delta $
to describe another possible control mechanism for optical pulses
in BGs \cite{Kashyap,Sterke}, namely a periodic \textit{chirp} of
the BG, i.e.~a local variation of the grating's period.

In this work we focus on the reflectivity superlattice, setting $\mu =0$;
the general situation including the periodic chirp will be considered
elsewhere. Thus $\epsilon$ and $k$ are the parameters of the model, which we
assume to be positive without loss of generality. According to the above
physical considerations, typically $\epsilon$ is small, but $k$ is not. Note
that we should only expect a quiescent GS to be stable in this kind of
nonuniform BG if the position of the soliton's center is located at a local
minimum of the reflectivity \cite{WMak}, i.e.~at $x=2\pi n/k$ for some
integer $n$. Hence we shall restrict our numerical search for GSs to just
such a case. Lastly, note that Eqs.~(\ref{e:uv}) conserve two dynamical
invariants, the Hamiltonian and norm (usually called \textit{energy} in
fiber optics), $E=\int_{-\infty }^{+\infty} \left(
|u(x)|^{2}+|v(x)|^{2}\right) dx$.

We shall seek stationary GS in the form $\left\{
u(x,t),v(x,t)\right\} =\exp (-i\omega t)\left\{
U(x),V(x)\right\}$, substitution of which into Eqs.~(\ref{e:uv})
yields
\begin{equation}
\begin{array}{rcl}
\omega U+iU^{\prime }+\left[ 1-\epsilon \cos (kx)\right] V+\left(
|V|^{2}+|U|^{2}/2\right) U & = & 0, \\
\omega V-iV^{\prime }+\left[ 1-\epsilon \cos (kx)\right] U+\left(
|U|^{2}+|V|^{2}/2\right) V & = & 0,\end{array} \label{e:uvans}
\end{equation}
where prime stands for $d/dx$. First, it is necessary to find
bandgaps in the system's spectrum. In the unperturbed problem
($\epsilon =0)$, the dispersion relation for linear waves with a
propagation constant $q$ is $\omega ^{2}=q^{2}+1$, thus producing
the well-known bandgap, $-1<\omega <1$. The spectrum for $\epsilon
>0$ should be found from the linearization of Eqs.
(\ref{e:uvans}). As in the Mathieu equation, gaps in these
equations emerge due to the parametric resonance caused by the
cosine modulation. Straightforward considerations show that new
bandgaps open up at points $\omega =\omega _{m}\equiv
\mathrm{sgn}(m)\sqrt{1+(mk)^{2}/4}$, with $m=\pm 1,\pm 2,\ldots $
(the extra modulation terms $\sim \mu $ in Eqs. (\ref{e:uv} ) open
gaps at the same positions). We will designate these gaps as
$\mathbf{\ m}^{\pm }$. Using perturbation theory, one can find
their boundaries for small $\epsilon $ and demonstrate that their
widths scale as $\epsilon ^{|m|} $. In particular, gaps
$\mathbf{1}^{\pm}$ are found in the region $|\chi |\leq 1$, $\chi
\equiv 2\left( \omega -\omega _{\pm 1}\right) /\epsilon $, to
leading-order in $\epsilon$.

\begin{figure}[t]
\begin{center}
\includegraphics[width=0.5 \linewidth]{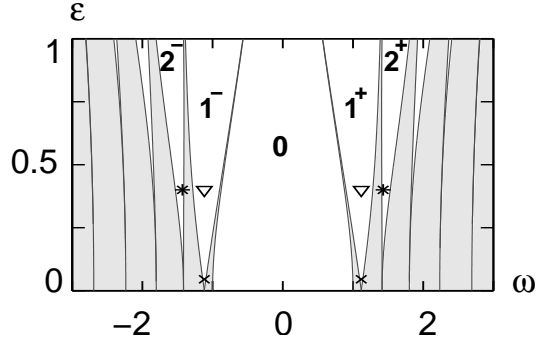}
\end{center}
\vspace*{-2ex}
\caption{The spectrum of Eq.~(\protect\ref{e:uv}) with $k=1$ in
the $(\protect\omega ,\protect\varepsilon )$ plane. Shaded and
white areas are, respectively, bands and gaps. Digits label the
gaps mentioned in the text. Markers correspond to the solutions
plotted in Fig.~\protect\ref{f:avdisc}.} \label{f:lingap}
\end{figure}

Equations~(\ref{e:uvans}) allow the invariant reduction $V=-U^{\ast }$,
which reduces them to a single equation,
\begin{equation}
\omega U+iU^{\prime }-\left[ 1-\epsilon \cos (kx)\right] U^{\ast
}+(3/2)|U|^{2}U=0  \label{e:unl}
\end{equation}
(the reduction with $V=U^{\ast }$ leads to an equation with
the opposite sign in front of $U^{\ast }$, which can be cast
back in the form of Eq. (\ref{e:unl}) by the substitution $U\equiv
i\tilde{U} $; neither reduction is valid for moving solitons, see
below). The linearization of Eq.~(\ref{e:unl}) is sufficient for
the analysis of the spectrum. Bandgap regions in the $(\omega
,\epsilon )$-parameter plane were computed using the software
package AUTO with driver HomMap \cite{Auto}. We detected points at
which the linearization of the Poincar\'{e} map \cite{GuHo:83}
around the origin ($U=0$) has eigenvalues $\pm 1$, and then
continued such points to identify the bandgap boundaries. The
results are displayed in Fig.~\ref{f:lingap}, which shows the
first five gaps. In this and subsequent examples, we set $k=1$, as
this value adequately represents the generic situation and is
physically meaningful for the application to fiber gratings.

In the central gap, which we designate $\mathbf{0}$, standard
calculations using Melnikov's method \cite{GuHo:83} reveal that
the GS solutions, which are known in an exact \ form for $\epsilon
=0$ \cite{AW-CJ:89}, extend to $\epsilon >0$. They were continued
numerically up to $\epsilon =1$, and were found to
\emph{completely fill} the central gap. That is, a GS exists for
each $\{\omega,\epsilon\}$-value belonging to the gap. We have
also computed eigenvalues that determine stability of these
solutions against small perturbations, and found that only GSs
with $\omega >0$ are stable. That is, the border between stable
and unstable GSs in the central gap for $\epsilon>0 $ is found to
remain, up to numerical accuracy, at $\omega =0$ as it is for
$\epsilon=0$ \cite{BPZ:98}.

Inside gaps $\mathbf{1^{\pm }}$, one can use the method of
averaging \cite{GuHo:83} to demonstrate the existence of GSs for
small $\epsilon $. To this end, we represent a solution to
Eq.~(\ref{e:unl}), $U(x)\equiv a(x)+ib(x)$, as
\begin{equation}
\left(
\begin{array}{c}
a \\
b\end{array} \right) =\sqrt{\epsilon }\exp \left( \left[
\begin{array}{cc}
0 & -(\omega _{\pm 1}+1) \\
\omega _{\pm 1}-1 & 0\end{array} \right] x\right) \left(
\begin{array}{c}
\Xi /k \\
\Theta /\left( 2\left( \omega _{\pm 1}+1\right) \right)\end{array}
\right) .  \label{e:trans}
\end{equation}
With constant amplitudes $\Xi $ and $\Theta $, we recover a
solution to the linearized equation (\ref{e:unl}) with $\epsilon
=0$ and $\omega =\omega _{\pm }$. The averaging method supposes
that $\Xi $ and $\Theta $ are functions of a slow coordinate,
$z\equiv x/\left[ 2k(\omega _{\pm 1}+1) \right] $, which leads to
equations
\begin{equation}
\frac{d}{dz}\left(
\begin{array}{c}
\Xi \\
\Theta\end{array} \right) =\left(
\begin{array}{c}
\alpha (1-\chi )\Theta -\beta \left( \Xi ^{2}+\Theta ^{2}\right) \Theta \\
\alpha (1+\chi )\Xi +\beta \left( \Xi ^{2}+\Theta ^{2}\right)
\Xi\end{array} \right) ,  \label{e:av}
\end{equation}
where $\alpha \equiv \epsilon \left[ k^{2}+4(\omega _{\pm
1}+1)^{2}\right]$, and $\beta \equiv 3\epsilon \lbrack
3k^{4}+8(\omega _{\pm 1}+1)^{2}k^{2}+48(\omega _{\pm
1}+1)^{4}]/[8k(\omega _{\pm 1}+1)]^{2}$. These equations conserve
their Hamiltonian, $H=\alpha \left[ (1+\chi )\Xi ^{2}-(1-\chi
)\Theta ^{2}\right] +(\beta /2)\left( \Xi ^{2}+\Theta ^{2}\right)
^{2}$. As gaps $\mathbf{1}^{\pm }$ exist for $|\chi |<1$, the
coefficients $1\mp \chi $ in Eqs.~(\ref{e:av}) and in $H$ are
positive. Therefore, the origin $(\Xi ,\Theta )=(0,0)$ is a saddle
in Eqs.~(\ref{e:av}), and a pair of homoclinic orbits to this
saddle can be found in the exact form
\begin{align}
(\tilde{\Xi}_{\pm }(t),\tilde{\Theta}_{\pm }(t))=& \left( \pm 2\sqrt{\frac{
\alpha }{3\beta }}\sin \theta \sin \frac{\theta }{2}\,\frac{\sinh (\alpha
t\sin \theta )}{\cosh (2\alpha t\sin \theta )+\cos \theta },\right. \\
& \qquad \left. \mp 2\sqrt{\frac{\alpha }{3\beta }}\sin \theta \cos \frac{
\theta }{2}\,\frac{\cosh (\alpha t\sin \theta )}{\cosh (2\alpha t\sin \theta
)+\cos \theta }\right) ,
\end{align}
with $\chi \equiv \cos \theta $. On application of the
transformation~(\ref{e:trans}), these orbits correspond to
solitons in the full system (\ref{e:unl}). In particular, there
are solutions with $a(x)=\mathrm{Re}\,U(x)$ odd and $b(x)=
\mathrm{Im}\,U(x)$ even. Thus, GSs exist in the entire gaps
$\mathbf{1^{\pm } }$ for sufficiently small $\epsilon $.

A similar analysis can be performed for higher bandgaps,
$\mathbf{2^{\pm }}$ , $\mathbf{3^{\pm }}$, etc., using a
higher-order averaging method \cite{Y-YK:99}. The respective
analytical computations show that each averaged system again
generates solitons. They have either even real and odd imaginary
parts or vice versa, depending on whether the gap's number $m$ is
odd or even.

\begin{figure}[tbp]
\begin{center}
\begin{minipage}{.4 \linewidth}
\includegraphics[width=0.8 \linewidth, height=0.52 \linewidth]
{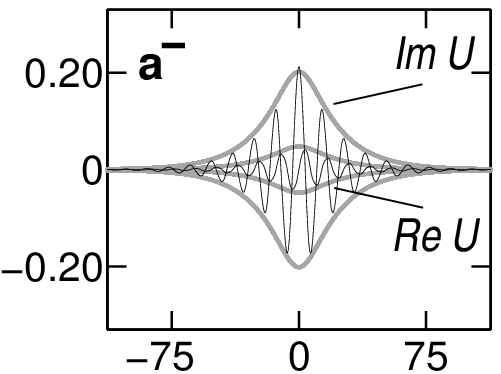}
\end{minipage}\hspace*{0 \linewidth}
\begin{minipage}{.4 \linewidth}
\includegraphics[width=0.8 \linewidth, height=0.52 \linewidth]{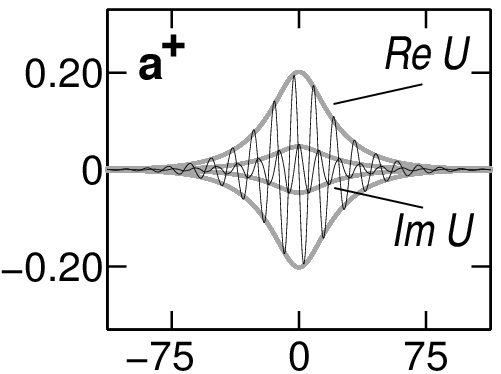}
\end{minipage}\\[1ex]
a) $\omega =\mp 1.118$, $\epsilon =0.045$, $k=1$ (points $\times$ in
Fig.~ \ref{f:lingap}) \\[2.5ex]
\begin{minipage}{.4 \linewidth}
\includegraphics[width=0.8 \linewidth, height=0.52
\linewidth]{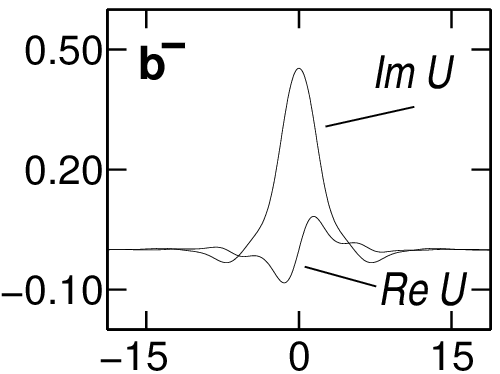}
\end{minipage}\hspace*{-0 \linewidth}
\begin{minipage}{.4 \linewidth}
\includegraphics[width=0.8 \linewidth, height=0.52 \linewidth]{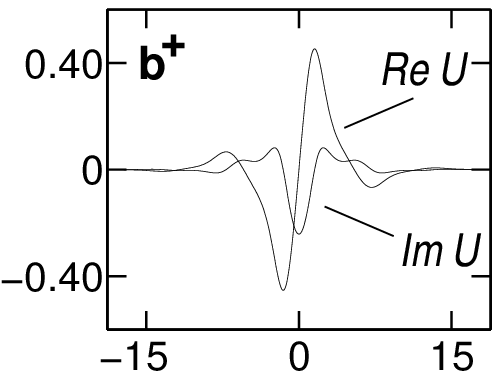}
\end{minipage}\\[1ex]
b) $\omega =\mp 1.118$, $\epsilon =0.4$, $k=1$ (points $\triangledown$ in
Fig.~\ref{f:lingap}) \\[2.5ex]
\begin{minipage}{.4 \linewidth}
\includegraphics[width=0.8 \linewidth, height=0.52
\linewidth]{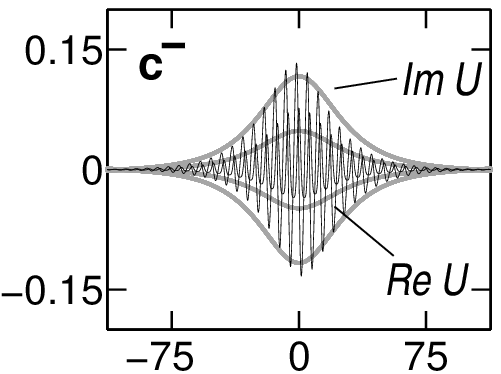}
\end{minipage}\hspace*{-0 \linewidth}
\begin{minipage}{.4 \linewidth}
\includegraphics[width=0.8 \linewidth, height=0.52 \linewidth]{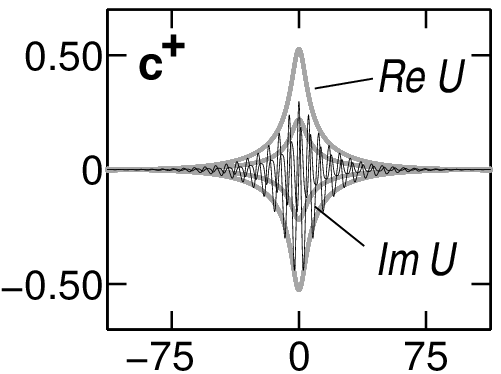}
\end{minipage}\\[1ex]
c) $\omega=\mp 1.42$, $\epsilon=0.4$, $k=1$ (points $\times \hspace*{-1.9ex}
+ $ in Fig.~\ref{f:lingap})
\end{center}
\caption{Stable fundamental soliton solutions of Eq.~(\protect\ref{e:unl})
in gaps $\mathbf{1^{\pm }}$ and $\mathbf{2}^{\pm }$. Gray lines in rows a)
and c) show solutions of averaged equations (\protect\ref{e:av}), and their
counterparts for gaps $\mathbf{2}^{\pm }$.}
\label{f:avdisc}
\end{figure}

\begin{figure}[t]
\begin{center}
\begin{minipage}{.4 \linewidth}
\includegraphics[width=.95 \linewidth, height=0.67
\linewidth]{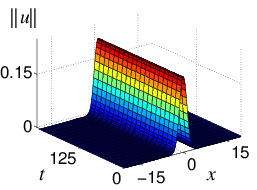} \\[-2ex]
(a) \hspace*{.8 \linewidth}
\end{minipage}
\hspace*{.1\linewidth}
\begin{minipage}{.4 \linewidth}
\includegraphics[width=.95 \linewidth, height=0.67
\linewidth]{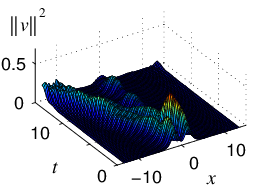} \\[-2ex]
(b) \hspace*{.8 \linewidth}
\end{minipage}
\end{center}
\caption{Soliton dynamics in gaps $\mathbf{1}^{\mp }$. (a)
Evolution of a perturbed symmetric soliton shown in Fig.
\protect\ref{f:avdisc}(b$^{-}$) ($\protect\epsilon =0.4$,
$\protect\omega =-1.118$; for these $\protect\epsilon $ and
$\protect\omega $, an asymmetric soliton exists too, but it is
unstable). (b) Splitting of the stable double-humped soliton shown
in Fig.~\protect\ref{f:avdisc}(b$^{+}$) into moving and quiescent
ones, after it was suddenly multiplied by $\exp \left( ipx\right)
$, with $p=1$.} \label{fig:stable}
\end{figure}

We employed the AUTO driver HomMap \cite{Auto} to continue
numerical soliton solutions of the full system~(\ref{e:unl}),
varying $\omega $ and $\epsilon $ . In so doing, a multitude of
symmetric and \emph{asymmetric} families of GS solutions,
including higher-order ones (bound states) were found in each new
gap. Here, we display numerical results for fundamental solitons,
as their bound states are likely to be unstable. It was found
that, as predicted above, each gap $\mathbf{1^{\pm }}$,
$\mathbf{2^{\pm }}$, etc. is completely filled with a single
family of symmetric GSs. The families in gaps $\mathbf{1}^{\pm }$
are represented by solitons displayed in
Figs.~\ref{f:avdisc}(a$^{\pm }$) and (b$^{\pm }$), for small and
larger $\epsilon $, respectively, while Figs.
\ref{f:avdisc}(c$^{\pm }$) displays the fundamental GSs in gaps
$\mathbf{2}^{\pm }$. Note that stable fundamental solitons may
feature a \emph{double-humped structure} in terms of $|U(x)|$ at
relatively large $\epsilon $, which is the case for the soliton in
Fig.~\ref{f:avdisc}(b$^{+}$ ), see Fig. \ref{fig:stable}(b) below
(however, the $|U(x)|$ profile of the GS displayed, for the same
$\epsilon $ but opposite $\omega $, in panel
\ref{f:avdisc}(b$^{-}$), remains single-humped, see Fig.
\ref{fig:stable}(a)). The homoclinic orbits of averaged system
(\ref{e:av}), also shown in a) and c), provide a good match to the
envelopes of the numerical solutions.

Recall that the ordinary GSs in gap $\mathbf{0}$ with $\epsilon
=0$ cannot be asymmetric, and they do not form bound states either
\cite{Sterke}. To explain where the asymmetric and higher-order
GSs for $\epsilon >0$ come from, we note that the new symmetric
fundamental GSs, found above in gaps $\mathbf{1}^{\pm }$,
correspond to transversal intersections of stable and unstable
manifolds in the Poincar\'{e} map associated with Eq.~(\ref{e:unl}
). Such intersections naturally occur in pairs, one representing a
symmetric soliton and the other one its asymmetric counterpart.
Moreover, the transversality of the intersection implies the
presence of a Smale horseshoe \cite{GuHo:83}, which, in turn,
ensures the existence of infinitely many higher-order GSs, that
may be realized as bound states of symmetric or asymmetric
fundamental solitons.

Stability of the GSs was tested by direct simulations of
Eqs.~(\ref{e:uv}), and verified through computation of the
eigenvalues of the equations linearized around a GS, which govern
the growth rate of small perturbations. Using this approach, it
was found that the \emph{entire} families of the symmetric
fundamental solitons that fill the gaps $\mathbf{1^{+}}$ and
$\mathbf{1^{-}}$ are stable. The significance of this result is
that we can have \emph{stable GSs\ }with $\omega <0$, something,
that is not possible in the ordinary BG model without the periodic
modulation terms. This finding is illustrated in
Fig.~\ref{fig:stable}(a), taking as an example the soliton with
$\omega=-1.118$ depicted in Fig.~\ref{f:avdisc}(b$^{-}$).

Regarding asymmetric solitons in gaps~$\mathbf{m^{\pm }}$ with $m\geq 1$, it
was found that some of them are stable and some are not, the unstable ones
being completely destroyed by growing perturbations (not shown here).

An important issue for practical application is the possibility of
finding moving GSs. (In experiments on ordinary fiber gratings,
only moving solitons have so far been observed \cite{experiment}).
In the presence of the superlattice, soliton mobility is a
nontrivial problem, because the GS has to move in a periodically
nonuniform medium. (Recall, though, that solitons belonging to
gaps $\mathbf{m}^{\pm }$ with $m\geq 1$ do not exist at all
without the superlattice). We have used numerical simulation to
test for mobility of the stable GSs we have so-far found.
Specifically, a quiescent stable soliton was multiplied by $\exp
(ipx)$, with $p>0$, which implies a sudden application of a
``shove" to the soliton, giving it momentum $P=i\int_{-\infty
}^{+\infty }\left( uu_{x}^{\ast }+vv_{x}^{\ast }\right)
dx+\mathrm{c.c.}\equiv 2pE$, where $E$ is the soliton's energy
defined above. It was observed that GSs belonging to central gap
$\mathbf{0}$ are readily set in stable motion by the shove. This
accords with the situation for $\epsilon =0$, where exact
solutions for moving GSs exist with any velocity $c $ from
$-1<c<+1$ \cite{AW-CJ:89}. In our simulations, we found that no
soliton in any gap could be made to move with velocity exceeding
$1$ (a very strong shove does not make the GS ``superluminal", but
simply destroys it).

For the stable GS belonging to gap $\mathbf{1}^{+}$, it was found
that the application of the shove caused the soliton to split into
a quiescent part and a moving part; see Fig.~\ref{fig:stable}(b).
A remarkable observation from these results is that the moving
soliton (which retains $\simeq 2/3$ of the initial energy) has a
negative average velocity, $c\approx -0.9$, hence its effective
mass $M\sim P/c$ is \emph{negative} too. (Simulations of the
moving GSs belonging to gap $\mathbf{0}$ have positive mass, as is
the case for $\epsilon =0$ \cite{WMak}). We remark that the mass
of a GS may be negative too in one- and two-dimensional models of
BEC in an ordinary optical lattice, and that this gives rise to
nontrivial effects such as stable confinement of solitons in an
\emph{anti-trapping} external potential \cite{HS}. Finally, the
shove applied to a stable GS belonging to gap $\mathbf{1}^{-}$ was
found to split into three solitons, one remaining quiescent while
the other two move off in opposite directions (not shown here).
Detailed results for moving solitons will be reported elsewhere.

In conclusion, we have investigated a model for a Bragg grating optical
fibre with a superlattice written on top. First, we identified a system of
new bandgaps in the fiber's spectrum, using an extension of the bandgap
theory for the Mathieu equation. Combining averaging methods and numerical
continuation, we have found that each new gap is completely filled with
fundamental symmetric solitons, which at some parameter values may have a
double-humped shape. In addition, new types of gap solitons (GSs) were found
that do not exist without the superlattice, such as asymmetric GSs and bound
states of fundamental GSs. An important finding is that the entire families
of fundamental GSs in the new bandgaps are stable, including negative
frequencies, where ordinary GSs are unstable. Stable moving solitons were
found too, including ones with a negative mass. Finally, it is pertinent to
point out that creation of the newly predicted GSs ought to be perfectly
feasible using presently available experimental techniques. In particular,
in a weak superlattice, with say, $\epsilon \simeq 0.1$, an estimate shows
that the solitons in the new bandgaps $\mathbf{1}^{\pm}$ can be observed if
the fiber grating, on top of which the superlattice is to be imposed, is of
length $\simeq 10$ cm.

\acknowledgments
K.Y. acknowledges support from the Japan Society for the Promotion of
Science for his stay at the University of Bristol. I.M.M. and B.A.M.
appreciate support from the Israel Science Foundation, through the
Center-of-Excellence grant No. 8006/03, and from EPSRC, UK. They thank the
Department of Engineering Mathematics at the University of Bristol for its
hospitality. T.W. acknowledges EPSRC support.

\end{document}